# Angular X-ray Cross-Correlation Analysis Applied to the Scattering Data in 3D Reciprocal Space from a Single Crystal

Dmitry Lapkin[1], Anatoly Shabalin[1], Janne-Mieke Meijer[2], Ruslan Kurta[3], Michael Sprung[1], Andrei V. Petukhov[4,5], Ivan A. Vartanyants[1,6]

[1]*Deutsches Elektronen-Synchrotron DESY, Notkestr. 85, 22607 Hamburg, Germany*

[2]*Department of Applied Physics & Institute for Complex Molecular Systems, Eindhoven University of Technology, 5600 MB Eindhoven, the Netherlands*

[3]*European XFEL, Holzkoppel 4, 22869 Schenefeld, Germany*

[4]*Debye Institute for Nanomaterials Science, Utrecht University, 3584 CS Utrecht, the Netherlands*

[5]*Laboratory of Physical Chemistry, Eindhoven University of Technology, 5612 AZ Eindhoven, the Netherlands*

[6]*National Research Nuclear University MEPhI (Moscow Engineering Physics Institute), Kashirskoe shosse 31, 115409 Moscow, Russia*

**February 17, 2022**

**Abstract**

We propose an application of the Angular X-ray Cross-Correlation Analysis (AXCCA) to the scattered intensity distribution measured in three-dimensional (3D) reciprocal space from a single crystalline sample. Contrary to the conventional application of AXCCA, when averaging over many two-dimensional (2D) diffraction patterns collected from different randomly oriented samples is required, the proposed approach gives an insight into the structure of a single specimen. This is particularly useful in studies of defect-reach samples that are unlikely to have the same structure. Here, we demonstrate an example of a qualitative structure determination of a colloidal crystal on the simulated as well as experimentally measured 3D scattered intensity distributions.

# 1. Introduction

The first approaches to study the structure of materials by means of the angular correlations in the scattered intensities go back to the late 70s – early 80s.[1-3] It was proposed by Zvi Kam to reveal the structure of macromolecules by analyzing the angular correlations in the scattering patterns from randomly oriented molecules in a solution.[1,2] In another research, correlations of scattered laser intensities from a colloidal glass were found to be related to its local structure.[3] At that time, the method did not undergo further development due to the lack of suitable instrumentation.[4] Recently, however, it became of great interest after the work of Wochner *et al.*,[5] where Angular X-ray Cross-Correlation Analysis (AXCCA) was applied to study the structure of colloidal glasses by means of X-ray scattering. The renewed interest to AXCCA was triggered by the development of modern X-ray sources such as synchrotrons of the 3$^{rd}$ and 4$^{th}$ generations[6] and novel X-ray Free Electron Lasers (XFELs)[7-10] that provide an X-ray beam with outstanding characteristics including high brilliance, ultimate coherence and femtosecond pulse durations. These characteristics allow measuring fluctuations in the scattering patterns that contain information about the local structure that could be revealed by AXCCA. The emergence of suitable equipment has led, among practical applications, to the development of the underlying theory.[11-16]

The practical applications of AXCCA are defined by the investigated sample and geometry of a typical X-ray scattering experiment. In such experiments, the scattered intensities are measured by a two-dimensional (2D) detector that represents a cut of reciprocal space by the Ewald sphere. AXCCA applied to such 2D patterns reveals symmetries of the sample in the plane orthogonal to the incident beam. This is particularly suitable in studies of (quasi-)2D samples such as 2D nanostructures[17-19], thin polymer films[20-22], liquid crystals[23-25]. In some cases, it is possible to refine the unit cell parameters of 3D superlattices of nanocrystals[26-29].

To explore the symmetries of a 3D sample, one typically collects many 2D patterns from randomly oriented identical samples, for example, injected bioparticles[30] or nanocrystals[36,31,32] as shown in Fig. 1(a). To achieve reasonable scattered intensities from a small single sample, extremely high flux of the incident X-ray beam is required, that can be provided by the modern XFELs. The diffraction patterns collected in such an experiment represent random cuts of reciprocal space, as shown in Fig. 1(b) that can be assembled into the intensity distribution in 3D reciprocal space. The main assumption of this approach is the reproducibility of the measured samples. If the measured samples are different, the revealed structure is averaged over many realizations.

At modern 3rd generation synchrotron sources one can use high coherence of these sources to reconstruct the 3D structure of the sample in Coherent Diffraction Imaging (CDI) experiments.[33,34] This may be achieved by the angular scan of the sample with a large unit cell in Small-Angle X-ray Scattering (SAXS) geometry as shown in Fig. 1(c-e). Finally, full reciprocal space of the sample is measured and is used for reconstruction. Although such reconstruction gives full information about the structure, the method is highly-demanding in terms of experimental requirements and data quality. AXCCA is based on analysis of angular correlations of scattered intensities in reciprocal space and can be applied to datasets of much lower quality, for which phase retrieval algorithms fail, to reveal the structural features averaged over the sample without the need to perform a reconstruction.

In this work, we propose to employ AXCCA to study symmetries of the intensity distribution in 3D reciprocal space from a single mesoscopic crystalline sample. We apply this method to simulated datasets for model colloidal structures and propose a geometrical model to interpret the results. As an example of practical application, we employ the dataset collected for a CDI reconstruction of a colloidal crystal grain.[33,35] We show that the developed method provides qualitative information about the real space structure without performing a complex iterative phase retrieval.

## 2. Theory

### 2.1. AXCCA applied to the intensity distribution in 3D reciprocal space

Here, we consider the scattered intensity distribution measured at different spatial positions by a 2D detector. The conventional AXCCA is based on the analysis of a two-point Cross-Correlation Function (CCF) defined as[31]

$$C(q_1, q_2, \Delta) = \langle \tilde{I}(\boldsymbol{q_1})\tilde{I}(\boldsymbol{q_2})\delta(\frac{\boldsymbol{q_1} \cdot \boldsymbol{q_2}}{\|\boldsymbol{q_1}\|\|\boldsymbol{q_2}\|} - \cos\Delta)\rangle, \quad (1)$$

where $\tilde{I}(\boldsymbol{q_1})$ and $\tilde{I}(\boldsymbol{q_2})$ are the scattered intensities measured by the detector at the points corresponding to the momentum transfer vectors $q_1$ and $q_2$ with the relative angle $\Delta$ between them. The averaging is performed over all positions corresponding to the momentum transfer vectors $q_1$ and $q_2$ with the lengths $q_1 = \|\boldsymbol{q_1}\|$ and $q_1 = \|\boldsymbol{q_2}\|$, respectively. The intensities can be scaled to their mean values, for example, as

$$\tilde{I}(\boldsymbol{q_i}) = \frac{I(\boldsymbol{q_i}) - \langle I(\boldsymbol{q_i})\rangle}{\langle I(\boldsymbol{q_i})\rangle}, \quad i = 1,2. \quad (2)$$

where averaging is performed over all measured intensities corresponding to the momentum transfer vectors $q_i$ with a certain length $q_i = \|q_i\|$.

When the measurements are performed in the small-angle X-ray scattering (SAXS) regime corresponding to small momentum transfer vectors, one can neglect the curvature of the Ewald sphere. Then, the definition in Eq. (1) simplifies to

$$C(q_1, q_2, \Delta) = \langle \tilde{I}(q_1, \varphi)\tilde{I}(q_2, \varphi + \Delta)\rangle_\varphi, \quad (3)$$

where $\tilde{I}(q,\varphi)$ is the scattered intensity measured by a detector at the position $\boldsymbol{q} = (q,\varphi)$, where $q, \varphi$ are the polar coordinates, and $\langle \cdots \rangle_\varphi$ denotes averaging over all angles $\varphi$.[11-16,23] The variables used in the definition of CCF (3) are shown in Fig. 2(a).

Typically, the CCFs are averaged over many 2D diffraction patterns collected from different realizations of the system (at different positions of the sample, at different times or from different randomly oriented injected particles). Averaging over many different system realizations allows suppressing random correlations in the scattered intensities specific to a certain realization of the system. The averaged CCFs represent the systematic correlations that correspond specifically to the internal structure of the samples and not to the certain realization of the system. Moreover, averaging over many orientations of the samples allows assessing the correlations in different cuts of 3D reciprocal space. Thus, the resulting CCFs represent all correlations in 3D reciprocal space and not only in certain planes.

In this work, we propose to apply Eq. (1) to the scattered intensity distribution in 3D reciprocal space measured for a single sample. In 3D reciprocal space, both momentum transfer vectors $\boldsymbol{q_1}$ and $\boldsymbol{q_2}$ can take any angular position. The averaging in Eq. (1) is then performed over spheres in reciprocal space with the radii $q_1=\|\boldsymbol{q_1}\|$ and $q_2=\|\boldsymbol{q_2}\|$, respectively, as shown in Fig. 2(b). The resulting CCFs in this case contain all present correlations from a single sample without a need to perform averaging over many realizations.

We note, a similar result would originate from averaging over many randomly oriented 2D scattering patterns collected from the same sample (or identical samples). Indeed, each pair of momentum transfer vectors taken in 3D reciprocal space lay in a certain 2D hyperplane that can be thought of as a 2D diffraction pattern. If the number of the randomly oriented 2D patterns is big enough, they cover the whole 3D space and the CCFs averaged over such a set of 2D patterns are identical to the CCFs calculated for the 3D pattern.[31] The number of

## 2.2. Cross-correlation functions in the case of a crystalline sample

AXCCA was shown to be useful to extract additional information from the scattering patterns from crystalline samples.[31,36] In this case, the scattered intensity contains well defined Bragg peaks originating from the crystallographic planes of the sample. When the CCF $C(q_1, q_2, \Delta)$ is calculated at the momentum transfer values $q_1$, $q_2$ corresponding to the Bragg peak positions, it contains correlation peaks at the characteristic relative angles $\Delta$ between the Bragg peaks, *i.e.* the reciprocal lattice vectors $g_1$ and $g_2$ with the lengths $q_1=\|g_1\|$ and $q_2=\|g_2\|$.

Given a model of a unit cell with the lattice basis vectors $a_1$, $a_2$ and $a_3$, one can calculate the reciprocal basis vectors $b_1$, $b_2$ and $b_3$ and thus any reciprocal lattice vector $\boldsymbol{g} = h\boldsymbol{b_1} + k\boldsymbol{b_2} + l\boldsymbol{b_3}$, $h, k, l \in \mathbb{Z}$.[37] For a pair of Bragg peaks corresponding to the reciprocal lattice vectors $g_1$ and $g_2$, the angle between them can be calculated using the scalar product

$$\boldsymbol{g_1} \cdot \boldsymbol{g_2} = \|\boldsymbol{g_1}\|\|\boldsymbol{g_2}\| \cos(\Delta). \qquad (4)$$

These Bragg peaks would contribute to the cross-correlation function calculated for the momentum transfer values $q_1$ and $q_2$ corresponding to the norms of the vectors $q_1 = \|g_1\|$ and $q_2 = \|g_2\|$, respectively, at the angle $\Delta$, as shown in Fig. 3(a). Given the lattice parameters and symmetry, one can calculate all positions of the correlation peaks. Details of such a calculation are given in Appendix A. We would like to note that in the case of high lattice symmetry, several pairs of different reciprocal lattice vectors with the same norms may contribute to the CCF at the same relative angle $\Delta$. For example, for a face-centered cubic (*fcc*) lattice, the pair of Bragg peaks 111 and $11\bar{1}$ as well as the pair 111 and $1\bar{1}1$ contribute to the CCF at the same angle $\Delta = \mathrm{acos}(1/3) \approx 70.53°$. In such a case, different peaks in the resulting CCFs can have different degeneracy, which is reflected in their relative magnitudes.

Considering close-packed structures, different stacking motifs of hexagonal layers result in different symmetries of the structures. Two structures of high symmetry are face-centered cubic (*fcc*) and hexagonal close-packed (*hcp*) lattices with the following stacking sequences: ABC for *fcc* and ABAB for *hcp*.[38] Stacking faults – irregularities in the stacking sequence – are very common defects in close-packed structures due to a low energy difference between the ideal structures.[39] A single inversion of the *fcc* stacking sequence ABCABCBACBA corresponds to a $\Sigma_3$-twinning boundary and results in two twinned *fcc* domains. Random stacking of hexagonal layers results in a so-called "random hcp" *rhcp* structure containing the motifs characteristic for both *fcc* and *hcp* structures. In reciprocal space, the stacking faults produce strong diffuse scattering in the stacking direction connecting the Bragg peaks in the form of rods that are known as Bragg rods, as shown in Fig. 3(b). Such Bragg rods are intensity modulations in reciprocal space along the straight lines connecting the Bragg peaks with fixed $h$ and $k$ indexes for which $h - k \neq 3n, n \in \mathbb{Z}$ and any index $l \in \mathbb{R}$ (in *hcp* notation). The Bragg peaks with indexes $h - k = 3n, n \in \mathbb{Z}$ and $l \in \mathbb{Z}$ are stacking-independent and are isolated in reciprocal space.[40] The intensity profiles along the Bragg rods depend on the particular stacking sequence as described in Ref.[35] In contrast to the isolated Bragg peaks that contribute to the CCFs at certain q-values, the Bragg rods contribute to the CFFs in a continuous *q*-range. Their contribution can be evaluated using the scalar product and corresponding reciprocal basis vectors as described in Appendix A.

3. Results

We demonstrate application of the AXCCA technique on simulated and experimentally measured datasets. The simulated datasets represent scattered intensity distributions in 3D reciprocal space calculated for colloidal crystal grains of different structures. The experimentally measured dataset is the scattered intensity distribution from a similar colloidal crystal that was studied previously.[33,35] Each of the datasets initially consisted of 360 diffraction patterns obtained by rotation of the sample in the range of 0 – 180° around the vertical axis with an angular step size of 0.5°. The simulation parameters were selected similar to those used in the experiment: the X-ray energy $E = 8$ keV ($\lambda = 1.55$ Å), the 2D detector (512 × 512 pixels) with the pixel size of 55 × 55 μm² positioned downstream from the sample at the distance of $d = 5.1$ m. The experimental dataset was collected at P10 Coherence Application beamline at PETRA III synchrotron using a MAXIPIX detector. The 2D patterns from each dataset were interpolated onto a 3D orthogonal grid with a voxel size of 0.4375 μm⁻¹. We used the flat Ewald sphere approximation because of small scattering angles (less than 0.25°, the corresponding $q$-values less than 200 μm⁻¹).

### 3.1. *Application to the Simulated Data*

For simulations, we considered a spherical colloidal crystal grain with the outer size of 3.6 μm consisting of monodisperse silica spheres with the diameter of 230 nm. Different close-packed structures typical for colloidal crystals were simulated: ideal *fcc* and *hcp* lattices, two *fcc* domains with a $\Sigma_3$-twinning boundary, as well as an *rhcp* lattice with the stacking sequence ABCABCBCBCACBCBABAB matching the one observed in the CDI reconstruction[33] of the experimental data discussed below. The nearest-neighbor distance for all the structures was equal to the diameter of the constituting silica spheres (230 nm). The simulated structures consist of corresponding stacking motifs of the hexagonal layers, as shown in Fig. 4(a,d,g,j).

The 2D diffraction patterns from the structures were simulated using MOLTRANS software. On the simulated diffraction patterns (see Fig. 4) one can observe rings of intensity due to the form factor of the colloidal spheres and the Bragg peaks that originate from the structure factor of the colloidal crystal lattice. In the diffraction patterns for the structures with the stacking faults (see Fig. 4(i,l)), besides the isolated Bragg peaks, the Bragg rods along the $q_z$-direction that connect Bragg peaks can be clearly seen.

The azimuthally averaged intensities of the 3D scattered intensity distributions for these structures are shown in Fig. 5(a). The intensity profiles for the ideal *fcc* and *hcp* lattices are quite different as they contain the characteristic Bragg peaks for these structures. In contrast, the profile for two twinned *fcc* domains with a $\Sigma_3$-boundary between them is almost identical to the one for the perfect *fcc* lattice. This is an expected result, because the major contribution to the scattered intensity originates from the domains with the same *fcc* structure, while the contribution from the boundary is negligible. The radial profile for the *rhcp* structure is smoothed and contain mostly the peaks common for the *fcc* and *hcp* structures that makes it hard to identify the exact stacking sequence. It is even harder in the case of the experimentally measured profile (shown in Fig. 5(a) for comparison) due to lower contrast.

We calculated the CCFs $C(q, \Delta) = C(q, q, \Delta)$ for the pairs of points with the same $q$ value $q = \|\boldsymbol{q}_1\| = \|\boldsymbol{q}_2\|$ in the simulated 3D intensity distributions for all four different structures (see Fig. 5(b) and Appendix B). We considered the CCFs for intensities at $q = 55$ μm$^{-1}$ that corresponds to stacking independent reflections present for all structures (see Fig. 5(a)). This $q$-value corresponds to the 220 reflections from the *fcc* structure and to the 110 reflections from the *hcp* structure. Even though these reflections correspond to the same *d*-spacing, the angles between the equivalent planes are different for these structures. Therefore, the peaks in the CCFs appear at different positions for different structures, as shown in Fig. 5(b). The peak positions from the geometrical model (see Appendix A) coincide with the peak positions in the calculated CCFs for the simulated structures as one can see from Fig. 5(b). The peak positions for an *fcc* structure are clearly distinct from those for an *hcp* structure because of different symmetry. The CCF for the twinned *fcc* structure contain additional peaks that are correlations between the peaks originating from different domains. The position of additional peaks is defined by the twinning transformation as described in Appendix A. This approach can be extended to other types of twinning (for example, $\Sigma_5$ or $\Sigma_9$) The CCF for the *rhcp* structure is

similar to the one for the twinned *fcc* domains, but the relative intensity of the peaks characteristic for the *hcp* structure to the ones characteristic uniquely for the twinned *fcc* structure is higher. This probably indicates the presence of both *hcp* and *fcc* stacking motifs, but more general conclusions can be made only by analyzing the CCFs calculated for different *q*-values as described below.

Additional information can be accessed if one looks at a set of CCFs calculated for various *q*-values. We calculated the CCFs $C(q,\Delta) = C(q,q,\Delta)$ in the range of $q = 25 - 115$ μm$^{-1}$ with a step size of 1 μm$^{-1}$ (see Fig. 6). As one can see from this figure, the peaks for the simulated structures have different positions in both radial and angular directions, since they originate from different sets of equivalent planes defined by the lattice symmetry. We would like to note that the negative values in the CCFs originate from subtraction of the mean intensity at each *q*-value according to Eq. (2). The resulting CCFs have zero mean value and, when peaks are present, the ground level has negative value. The peak positions for these structures can be calculated from the geometrical model of the reciprocal lattice as described in Appendix A. We note that the peak positions were determined for the structures with the unit cell parameters corresponding to the nearest neighbor distance of 230 nm (the size of the silica spheres). In an arbitrary experiment, the unit cell parameters can be used as the fitting parameters to fit the peak positions in the experimental CCFs.[34,41]

For the ideal *fcc* and *hcp* structures, the positions of all brightest peaks in the CCFs coincide with the positions obtained from the geometrical model (see Fig. 6(a,b)). Additionally, there are low-intensity peaks at the *q*-values between the bright peaks that are not explained with this model (see, for example, additional peaks at q = 36 μm$^{-1}$ in Fig. 6(a)). They originate from the correlations between the Bragg peaks of different orders. Basically, different orders contribute to the scattered intensities at different *q*-values, but due to the broadening of the Bragg peaks and the absence of noise in the simulated data, their tails contribute to the CCFs.

They are not observed in the experimental data due to noise and other artefacts, but they also can be considered in the simple geometrical model.

For the twinned *fcc* structure, the map contains many additional peaks that reflect correlations between the Bragg peaks originated from different domains. As discussed above, the peak positions are defined by the twinning transformation and can be taken into account as described in Appendix A. For the *rhcp* structure, the map contains peaks characteristic for both *hcp* and *fcc* structures. It is rather an expected result as soon as the *rhcp* structure contains stacking sequences that can be attributed to both *hcp* and *fcc* structures. Besides the peaks, the CCFs for the *rhcp* and twinned-*fcc* structures contain also intensity in the form of "arcs" connecting the peaks. They originate from the Bragg rods characteristic for stacking disordered structures with planar defects. Their contribution to the CCFs can be calculated as described in Appendix A and shown in Fig. 6(d).

Despite similar intensity profiles, different structures result in different angular distribution of the Bragg peaks. The AXCCA technique allows one to reveal the angular correlations between the Bragg peaks and to determine qualitatively the sample structure even when the azimuthally integrated intensity profiles are almost identical.

### 3.2. *Application to the Experimental Data*

The experimentally measured sample was a colloidal crystal grain with an outer size of about $2 \times 3 \times 4$ μm$^3$ consisted of polystyrene spheres with the diameter of about 230 nm prepared as described in Refs.[33,35] The collected scattered intensity distribution in 3D reciprocal space contains several orders of Bragg peaks and Bragg rods (see Fig. 7(a)). An in-plane cut through the origin of reciprocal space (see Fig. 7(b)) reveals the 6-fold symmetry characteristic for hexagonal layers of close-packed nanoparticles. Two out-of-plane cuts shown in Fig. 7(c,d) contain the Bragg rods connecting the Bragg peaks indicating the stacking disorder of the nanoparticle layers. It should be noted that the experimentally measured diffraction patterns have significantly lower contrast in comparison to the simulated ones. This can be attributed to the polydispersity of the colloidal particles, the partial coherence of the incident X-rays and other experimental artifacts that are not taken into account in the simulations.

The experimental CCFs $C(q,\Delta) = C(q,q,\Delta)$ calculated for the pair of points with the same $q$ values $q = \|\boldsymbol{q}_1\| = \|\boldsymbol{q}_2\|$ in the range of $q$ = 25 – 115 μm$^{-1}$ with a step size of 1 μm$^{-1}$ are shown in Fig. 8(a). Due to lower contrast of the diffraction patterns, these correlation maps also have lower contrast in comparison to the simulated ones. Moreover, the measured intensity in the locations of form factor minima does not contain any structural information leading to the absence of the peaks in the CCFs at the corresponding $q$-values. We assumed that the colloidal crystal has a close-packed structure and calculated the peak positions in the CCFs according to the geometrical model for the same structures as for the simulated data: ideal *fcc*, *hcp* and twinned *fcc*. Also, we calculated the positions of the "arcs" corresponding to the correlations between the Bragg rods. The experimental CCFs with the indicated peak positions are shown in Fig. 8(b-d).

Most of the peaks present in the experimental CCFs have the peak positions characteristic for an *hcp* structure (see Fig. 8(c)) indicating that this stacking motif is a predominant one. Several peaks do not match the positions for the *hcp* structure, but their positions are characteristic for an *fcc* structure (see Fig. 8(b)) suggesting presence of such stacking motif in the sample as well. The peaks characteristic for twinned *fcc* domains are not present in the experimental CCFs (compare with Fig. 6(c)), that indicates absence of such motifs in the sample. In addition, there are "arcs" characteristic for correlations between the Bragg rods similar to the ones for the simulated *rhcp* structure.

Indeed, the stacking sequence revealed in the reconstructed real space structure is ABCABCBCBCACBCBABAB.[33] This sequence, in general, can be described as random *hcp* structure with many stacking faults. However, one can distinguish *hcp* and *fcc* motifs in the sequence that results in the corresponding peaks in the CCFs.

### *3.3. Comparison of AXCCA applied to the intensity distribution in 3D reciprocal space and to the randomly oriented 2D diffraction patterns*

As mentioned in the section 2, the CCFs calculated for the intensity distribution in 3D reciprocal space should be similar to the ones averaged over many 2D diffraction patterns obtained from different random angular orientations of the same sample. Such a dataset of 2D diffractions patterns could be collected in an XFEL experiment performed in the single particle imaging (SPI) experiment, if the same crystalline structure was injected into the X-ray beam many times in random orientations. To prove the identity of the CCFs obtained from the 3D intensity distribution and the ones averaged many 2D diffractions patterns in random orientations, we simulated $5 \cdot 10^4$ diffraction patterns from the randomly oriented colloidal crystal with the *fcc* structure using the MOLTRANS software as described in Section 3.1. The angular orientations were uniformly distributed in 3D. The CCFs $C(q,\Delta) = C(q,q,\Delta)$ were calculated for *q*-values in the range of $q = 25 – 115$ μm$^{-1}$ using Eq. (3) for each diffraction pattern separately and then averaged over all patterns.

The resulting CCFs averaged over all $5 \cdot 10^4$ patterns are shown in Fig. 9(b) and can be compared to the ones calculated for the intensity distribution in 3D reciprocal space as described in Section 3.1 and shown in Fig. 9(a). As one can see from these figures, the CCF maps are almost identical and contain peaks at the same positions. Small deviations probably originate from interpolation of the scattered intensity onto the 3D grid in the second case.

In contrast, the CCF maps averaged over 5·10² 2D patterns, shown in Fig. 9(c), contain only a fraction of the peaks present in the CCF map calculated for the 3D intensity distribution. This is because such a small number of patterns do not fully cover all possible orientations. Indeed, to contribute into the CCF, a pair of Bragg peaks should be present in a single 2D diffractions pattern. Thus, it requires a certain number of randomly oriented diffractions patterns to catch all possible pairs of the Bragg peaks.

To estimate the number of 2D diffraction patterns in random orientations required to obtain a CCF map similar to the one calculated from the 3D scattered intensity distribution, we calculated the Pearson correlation coefficient[42] $r(N)$ between the CCF maps averaged over different numbers of 2D patterns and the one from the 3D intensity distribution defined as

$$r(N) = \frac{\langle C_{3D}(q,\Delta) C_N(q,\Delta) \rangle}{\langle C_{3D}^2(q,\Delta) \rangle \langle C_N^2(q,\Delta) \rangle}, \qquad (5)$$

where $C_{3D}(q,\Delta)$ are the CCFs calculated for the intensity distribution in 3D reciprocal space, and $C_N(q,\Delta)$ are the CCFs calculated for the randomly oriented 2D diffraction patterns and averaged over $N$ of them. The averaging was performed over all $q$-values in the range of $q = 25 - 115$ μm$^{-1}$ and angles $\Delta = 0 - 180°$ for which the CCFs were calculated. We would like to note that the calculated CCFs have zero mean value with averaging over angle $\Delta$ at the fixed $q$-value that allows direct application of the Pearson correlation coefficient.

The evolution of the correlation coefficient with the number of diffraction patterns is shown in Fig. 9(d). When the number of patterns used is below $10^2$, the correlation coefficient is close to zero indicating that the CCFs do not contain any features corresponding to the structural information. With further increase in the number of patterns used, the correlation coefficient grows that indicate the successive appearance of the structured features in the CCF map. At about 3·10³ patterns it reaches a plateau, while with further increase in the number of patterns it grows only a little bit to the value of 0.95 for 5·10⁴ patterns. We suggest that all features in the CCF map appear already at $3 \cdot 10^3$ patterns, while further increase in the number of patterns lead to only minor changes in the relative intensities of the correlation peaks.

Thus, the CCFs calculated from the 2D diffractions patterns obtained for different random orientations of the sample are similar to the CCFs calculated from the scattered intensity distribution in 3D reciprocal space measured for the sample, when the number of 2D patterns is high enough. In the particular case under consideration, the number of required randomly oriented 2D diffraction patterns is about two orders of magnitude higher than the number of systematically measured 2D patterns (for example, by rotation of the sample) required for reconstruction of the intensity distribution in 3D reciprocal space.

We note that the number of randomly oriented 2D diffraction patterns required to obtain the CCF map similar to the one calculated from the 3D intensity distribution is individual for each sample under study. The number of required patterns depends on the probability to catch at least a pair of the Bragg peaks into a single 2D pattern that in turn depends on the angular size and separation of the Bragg peaks in 3D reciprocal space. Therefore, for bulk crystals with many scatterers and small periodicity, the required number of 2D patterns may be sufficiently higher.

The important point here is the distribution of the angular orientations of the sample, for which 2D diffraction patterns are obtained. Only uniform angular distribution allows obtaining the CCFs similar to those obtained from the 3D intensity distribution, because the 2D patterns in this case cover all pairs of the points in reciprocal space with equal probability. If the angular distribution is not uniform, some correlations will be enhanced, while others – weakened.

To show that, we simulated 2D diffraction patterns from a colloidal crystal with the *fcc* structure using the MOLTRANS software as described in Section 3.1. We simulated two datasets, obtained by rotation of the sample around the $[111]_{fcc}$ and $[1\bar{1}0]_{fcc}$ axes in the range of 0 – 180° with an angular step size of 0.5°. The CCFs were calculated for the 2D diffraction patterns in the *q*-range of 25 – 115 μm$^{-1}$ with the step size of 1 μm$^{-1}$ and then averaged over all angular positions. The resulting CCFs calculated for the two datasets are different between each other and from our initial map shown in Fig. 6(a), as it is well seen in Fig. 9(e-f). The difference can be explained as follows. A pair of the Bragg peaks gives rise to a peak in the CCFs only if both Bragg peaks are present in the same 2D diffraction pattern. Moreover, the intensities of the present correlation peaks in this case are enhanced in comparison to the ones obtained for the randomly distributed 2D patterns. This is because, in the latter case, the CCFs are averaged over many patterns, most of which do not contain any correlations at a certain *q*-value. The diffractions patterns obtained by the rotation around one crystallographic axis are an extreme case, but any other distribution with a preferred direction would result in similar deviations.

**Conclusions**

We proposed to apply the AXCCA technique to the scattered intensity distribution in 3D reciprocal space. Here, we demonstrated an application of the AXCCA for qualitative determination of the crystalline structure of a colloidal crystal, including the present planar defects. AXCCA provides a complementary view on the structure when CDI reconstruction does not work.[41] The results can be interpreted by means of a simple geometrical model of the crystalline lattice and defects. Direct sensitivity to the angles in reciprocal space provides additional information about the structure in comparison to the conventional radial intensity profile analysis.

The application of AXCCA to the 3D scattered intensity distribution measured from a single sample by its rotation made it possible to avoid averaging of the revealed structure over many realizations with possibly different defects present. Moreover, the systematic measurement allowed to significantly reduce the number of measurements to obtain orientationally averaged CCFs, as compared to measurements from random orientations. We also showed that averaging over 2D diffraction patterns measured while rotation around the fixed axis does not provide the same CCFs as the assembly of intensity distributions in 3D reciprocal space. We think that it is an essential part of the proposed method.

The method described here works well for the colloidal samples with the large unit cell. For such samples rotation over one axis is sufficient to obtain information about the whole reciprocal space. The same method can be applied as well for the crystal grains with the unit cell of few angstroms. In this case due to Ewald sphere curvature one will need to apply two rotations around two orthogonal axes to cover full reciprocal space of the crystal grain. Described in this work formalism will be applicable also in this case.

This approach was already successfully applied for the analysis of the averaged structures and defects in single grains of Au and magnetite.[34,41] We expect that it will find a lot of applications for understanding the structure of colloidal grains and single crystals in future.

**Acknowledgements**

The authors are thankful to E. Weckert for supporting the project and providing the MOLTRANS software for simulating the scattering patterns. I. A. V. acknowledges the financial support of the Russian Federation represented by the Ministry of Science and Higher Education of the Russian Federation (Agreement No. 075-15-2021-1352).

*APPENDIX A*

*Geometrical interpretation of the CCFs*

Here, we follow discussion provided in Ref.[31] and apply it to our structures. Any reciprocal lattice vector can be represented as a linear combination of the basis vectors $\boldsymbol{g}_{hkl} = h\boldsymbol{b}_1 + k\boldsymbol{b}_2 + l\boldsymbol{b}_3$. Let us denote a family of equivalent crystallographic directions as $G_{hkl}$. Each crystallographic direction that fulfills the diffraction selection rules for a given lattice symmetry corresponds to the position of a Bragg peak in reciprocal space. Using the coordinates of the reciprocal lattice vectors *g*$_{hkl}$ and *g*$_{h'k'l'}$, one can calculate the angle Δ between a certain pair of the Bragg peaks corresponding to these vectors. This Bragg peaks pair would contribute at the angle Δ to the CCF *C*(*q$_1$*,*q$_2$*,Δ) calculated for $q_1 = \|\boldsymbol{g}_{hkl}\|$ and $q_2 = \|\boldsymbol{g}_{h'k'l'}\|$. To evaluate all contributions for a certain fixed *q$_1$* and *q$_2$*, one should consider all Bragg peaks that appear in reciprocal space at these *q*-values. Then, all angles Δ can be calculated using the scalar product

$$\boldsymbol{g}_{hkl} \cdot \boldsymbol{g}_{h'k'l'} = \|\boldsymbol{g}_{hkl}\|\|\boldsymbol{g}_{h'k'l'}\| \cos(\Delta), \quad (7)$$

if one considers all possible pairs of the vectors $\boldsymbol{g}_{hkl}$ and $\boldsymbol{g}_{h'k'l'}$ from certain families of equivalent crystallographic directions $G_{hkl}$ and $G_{h'k'l'}$, respectively, corresponding to the *q*-values *q$_1$* and *q$_2$*. One should note that, in some cases, several families $G_{h_i k_i l_i}$ may contribute at the same q-value. Then, an extended set of the vectors $\cup_i G_{h_i k_i l_i}$ should be considered.

Several crystalline domains in the sample would results in two types of the correlations: intra-domain correlations between the Bragg peaks originating from a single domain and inter-domain correlations between the Bragg peaks originating from different domains. The intra-domain correlations contribution into the CCFs can be evaluated as described above. To evaluate the contribution of the inter-domain correlations, one should consider the relative orientation of the domains. The orientation can be taken into account by introducing an orthogonal transformation matrix **T** that transform the basis vectors of one domain into the basis vectors of another one. Then, the inter-domain correlations contribute to the CCF $C(q_1,q_2,\Delta)$ at the angles $\Delta$ that can be found using the scalar product

$$\boldsymbol{g_{hkl}} \cdot \boldsymbol{T g_{h'k'l'}} = \|\boldsymbol{g_{hkl}}\|\|\boldsymbol{g_{h'k'l'}}\| \cos(\Delta), \tag{8}$$

if one considers all possible pairs of the vectors $\boldsymbol{g_{hkl}}$ and $\boldsymbol{g_{h'k'l'}}$ from certain families of equivalent crystallographic directions $G_{hkl}$ and $G_{h'k'l'}$ corresponding to the $q$-values $q_1$ and $q_2$.

For example, for *fcc* and *hcp* lattices, discussed in this paper, the reciprocal basis vectors can be defined as follows:

$$\begin{cases} b_1^{fcc} = \dfrac{2\pi}{a_{fcc}}\left(\dfrac{1}{\sqrt{2}}, -\dfrac{1}{\sqrt{6}}, \dfrac{1}{\sqrt{3}}\right) \\ b_2^{fcc} = \dfrac{2\pi}{a_{fcc}}\left(-\dfrac{1}{\sqrt{2}}, -\dfrac{1}{\sqrt{6}}, \dfrac{1}{\sqrt{3}}\right) \\ b_3^{fcc} = \dfrac{2\pi}{a_{fcc}}\left(0, \dfrac{2}{\sqrt{6}}, \dfrac{1}{\sqrt{3}}\right) \end{cases} \text{and} \begin{cases} b_1^{hcp} = \dfrac{2\pi}{a_{hcp}}\left(\dfrac{2}{\sqrt{3}}, 0, 0\right) \\ b_2^{hcp} = \dfrac{2\pi}{a_{hcp}}\left(\dfrac{1}{\sqrt{3}}, 1, 0\right) \\ b_3^{hcp} = \dfrac{2\pi}{a_{hcp}}\left(0, 0, \dfrac{\sqrt{3}}{\sqrt{2}}\right) \end{cases} \tag{9}$$

where $a_{fcc} = \sqrt{2}d$ and $a_{hcp} = d$ are the *fcc* and *hcp* lattice parameters corresponding to the same nearest-neighbor distance $d$. The orientation of the *fcc* basis is selected in such a way that the stacking directions $[001]_{hcp}/[111]_{fcc}$ coincide as well as the angular orientation of the hexagonal planes $(001)_{hcp}/(111)_{fcc}$ (i.e. $[100]_{hcp} \| [1\bar{1}0]_{fcc}$).

For the discussed in this paper simplest *fcc* twinning with a $\Sigma_3$-boundary, the transformation matrix **T** corresponds to a reflection of the *fcc* lattice across the $(111)_{fcc}$ plane and can be written in the following form:

$$\boldsymbol{T} = \begin{pmatrix} 1 & 0 & 0 \\ 0 & 1 & 0 \\ 0 & 0 & -1 \end{pmatrix}. \tag{10}$$

The Bragg rods originating from the stacking disorder of the hexagonal layers in close-packed structures are intensity modulations along the straight lines normal to the hexagonal layers and along the stacking direction. Their positions are defined by the reciprocal lattice and, using the *hcp* reciprocal basis vectors described in Eq. (9), can be described as $\boldsymbol{g_{hk}}(l) =$ $\mathbf{g}_{hk0} + \mathbf{g}_\perp(l)$, where $\mathbf{g}_{hk0}$ is a vector from a certain in-plane Bragg peaks family $G_{hk0}^{hcp}$ and $\mathbf{g}_\perp(l) = l\mathbf{b}_3^{hcp}, l \in (-\infty, \infty)$ is a vector along the Bragg rod, normal to the planes. One should note that the Bragg rods are present only for stacking-dependent families $G_{hk0}^{hcp}$ for which $h - l \neq 3n, n \in \mathbb{Z}$.[30]

The parameter $l$ corresponding to a certain $q$-value can be easily calculated for any Bragg rod corresponding to a certain in-plane reciprocal lattice vector $\mathbf{g}_{hk0}$ as

$$l = \frac{\sqrt{q^2 - \|\mathbf{g}_{hk0}\|^2}}{\|\mathbf{b}_3\|}. \tag{11}$$

Then, a pair of Bragg rods, corresponding to different in-plane vectors $\mathbf{g}_{hk0}$ and $\mathbf{g}_{\|2}$, contributes to the CCF $C(q_1,q_2,\Delta)$ at the angle $\Delta$ that can be calculated using the scalar product

$$\mathbf{g}_{hk}(l_1) \cdot \mathbf{g}_{h'k'}(l_2) = \|\mathbf{g}_{hk}(l_1)\| \|\mathbf{g}_{h'k'}(l_2)\| \cos(\Delta), \tag{12}$$

where $\mathbf{g}_{hk}(l_i) = \mathbf{g}_{hk0} + \mathbf{g}_\perp(l_i)$, $\mathbf{g}_{hk0}$ is a vector of a certain family $G_{hk0}^{hcp}$, $\mathbf{g}_\perp(l_i) = l_i \mathbf{b}_3^{hcp}$ and parameter $l_i$ corresponding to the $q$-value $q_i$ is defined by Eq. (11).

## APPENDIX B

### Definition of the Cross-Correlation Function for the intensities defined on a grid

Taking into account that the experimental data are typically defined on a grid in reciprocal space, Eq. (2) can be presented as

$$C(q_1, q_2, \Delta) = \frac{\sum_{\{|\|q_i\|-q_1|<\varepsilon\} \cap \{|\|q_j\|-q_2|<\varepsilon\} \cap \{\Delta_{ij} \in [\Delta-d\Delta;\Delta+d\Delta]\}} \tilde{I}(q_i)\tilde{I}(q_j)}{\sum_{\{|\|q_i\|-q_1|<\varepsilon\} \cap \{|\|q_j\|-q_2|<\varepsilon\} \cap \{\Delta_{ij} \in [\Delta-d\Delta;\Delta+d\Delta]\}} 1}, \quad (5)$$

where $q_i$, $q_j$ are the points close to the spheres of the radii $q_1$ and $q_2$ in reciprocal space, respectively, $\Delta_{ij}$ is the relative angle between these points. The sum is calculated over all pairs of points $q_i$, $q_j$ with the corresponding relative angle $\Delta$. The average intensity used for the intensity correction in this case is

$$\langle I(q) \rangle = \frac{\sum_{|\|q_i\|-q|<\varepsilon} I(q_i)}{\sum_{|\|q_i\|-q|<\varepsilon} 1} \quad (6)$$

Given the desired resolution of 1 μm$^{-1}$, in this work the radial averaging window $\varepsilon$ was selected to be 0.5 μm$^{-1}$. The angular resolution of $\Delta$ was experimentally set to 0.5° that allows one to resolve all peaks in the resulting CCFs. The angular averaging window $d\Delta$ was correspondingly set to 0.25°.

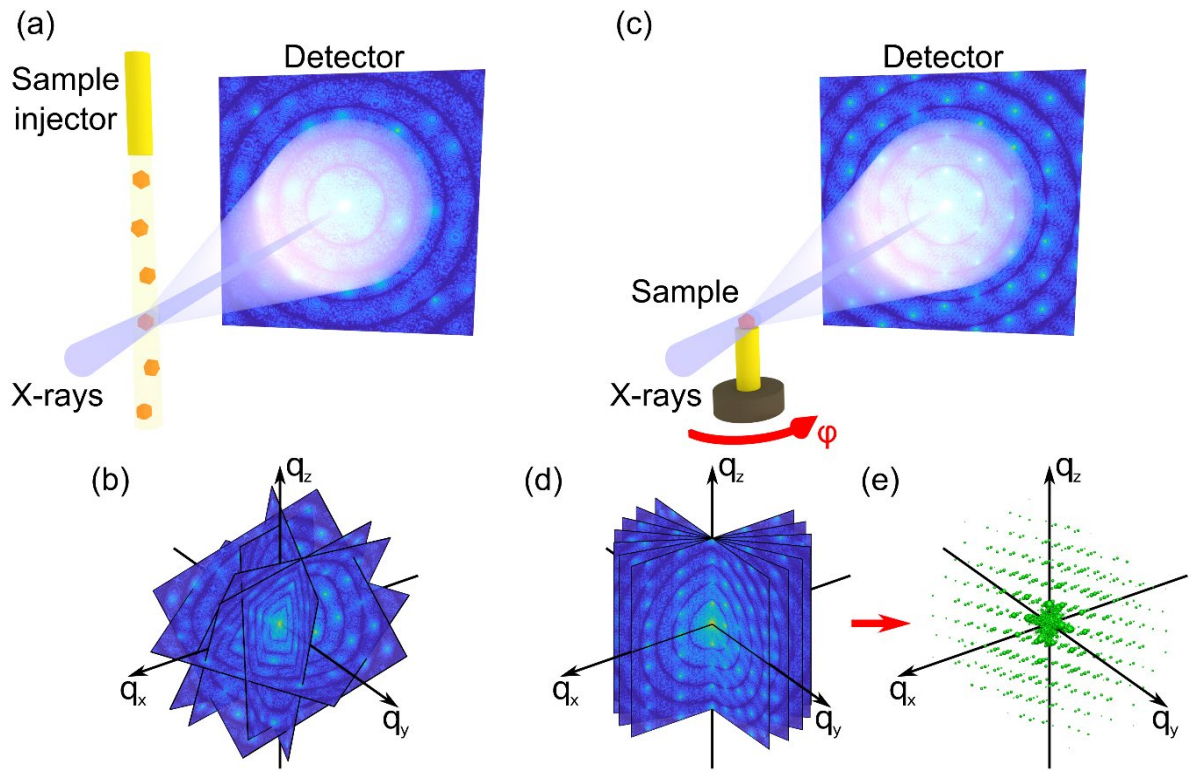

**Figure 1. (a)** Possible scheme of the experiment setup for measuring 2D diffraction patterns from different randomly oriented samples injected into the incident X-ray beam. The collected this way patterns represent random cuts of the 3D reciprocal space as shown in **(b)**. **(c)** Possible scheme of the experiment setup for measuring 2D diffraction patterns from a single sample rotated around an axis normal to the incident beam. The 2D patterns of known orientation **(d)** can be further interpolated into 3D intensity distribution **(e)**.

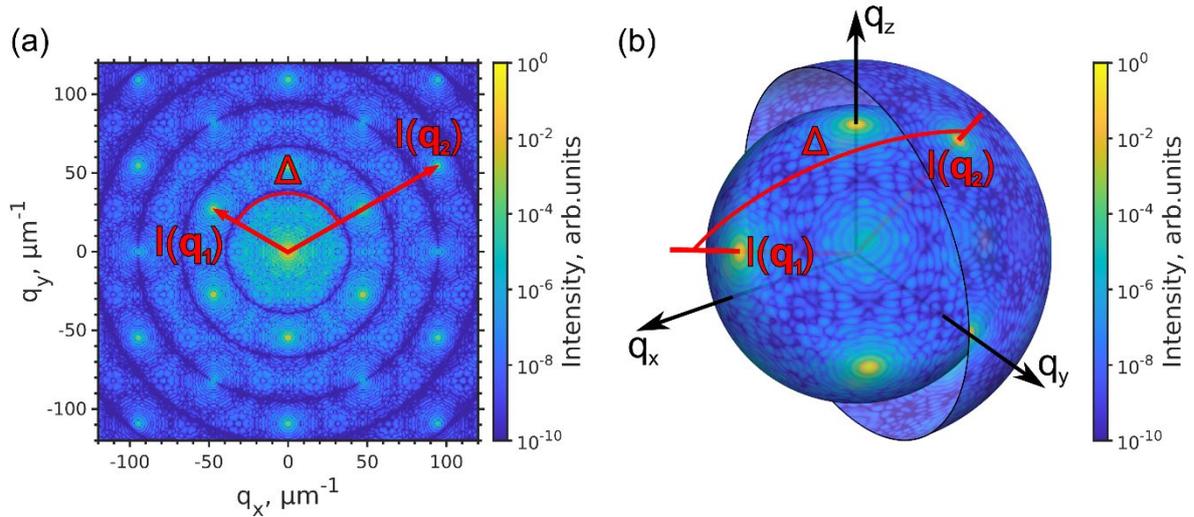

**Figure 2.** Scheme of the CCF calculation in the case of **(a)** 2D and **(b)** 3D intensity distributions. The product of intensities at two points $q_1$ and $q_2$ in reciprocal space, separated by the angle $\Delta$, contribute into the CCF value at this $\Delta$ value. The final CCF is obtained by averaging over all points on the rings/spheres of the corresponding radii. The color code exemplarily represents the simulated intensities for a colloidal crystal with *fcc* structure: (a) 2D diffraction pattern from the colloidal crystal oriented along [001] direction in respect to the incident X-ray beam and (b) intensities at the spheres in the 3D reciprocal space of the colloidal crystal with the radii $q_1$ and $q_2$, corresponding to 111 and 220 reflections, respectively.

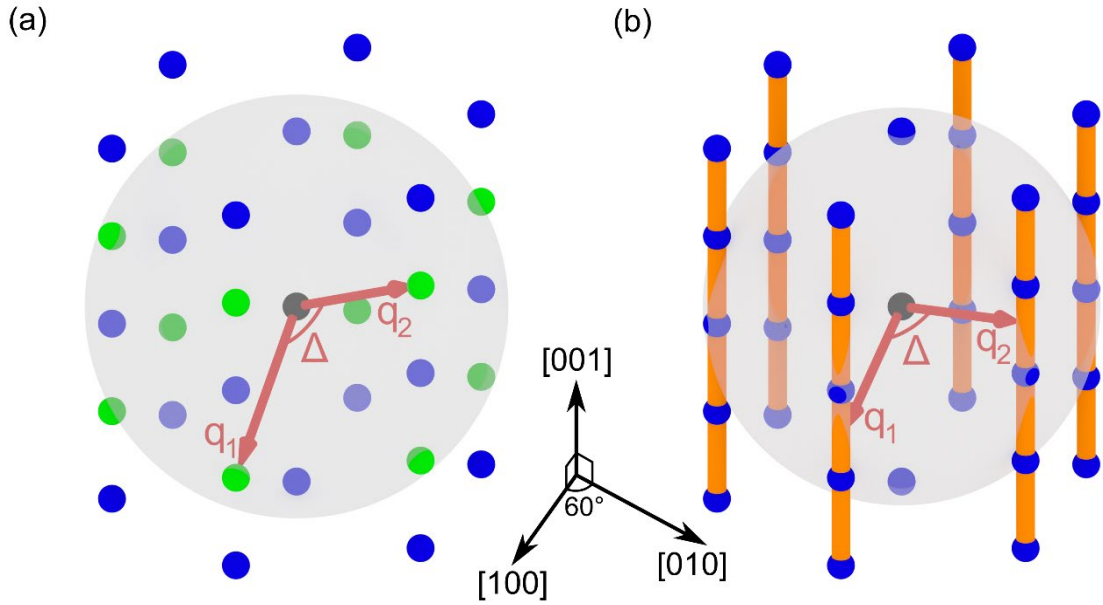

**Figure 3.** Models of 3D reciprocal space for an **(a)** *hcp* and **(b)** *rhcp* lattices. The black sphere is the origin of reciprocal space, colored spheres – the Bragg spots. The semitransparent sphere shows the sphere $S$ of the radius $q = \|q_1\| = \|q_2\|$, at which the CCF is calculated. In (a), the green spheres are the Bragg spots intersecting the sphere $S$ and, thus, contributing to the corresponding CCF at the angle $\Delta$. In (b), the orange rods represent the Bragg rods. They contribute to the corresponding CCF at the angle $\Delta$ that is dependent on the radius $q$ of the sphere $S$.

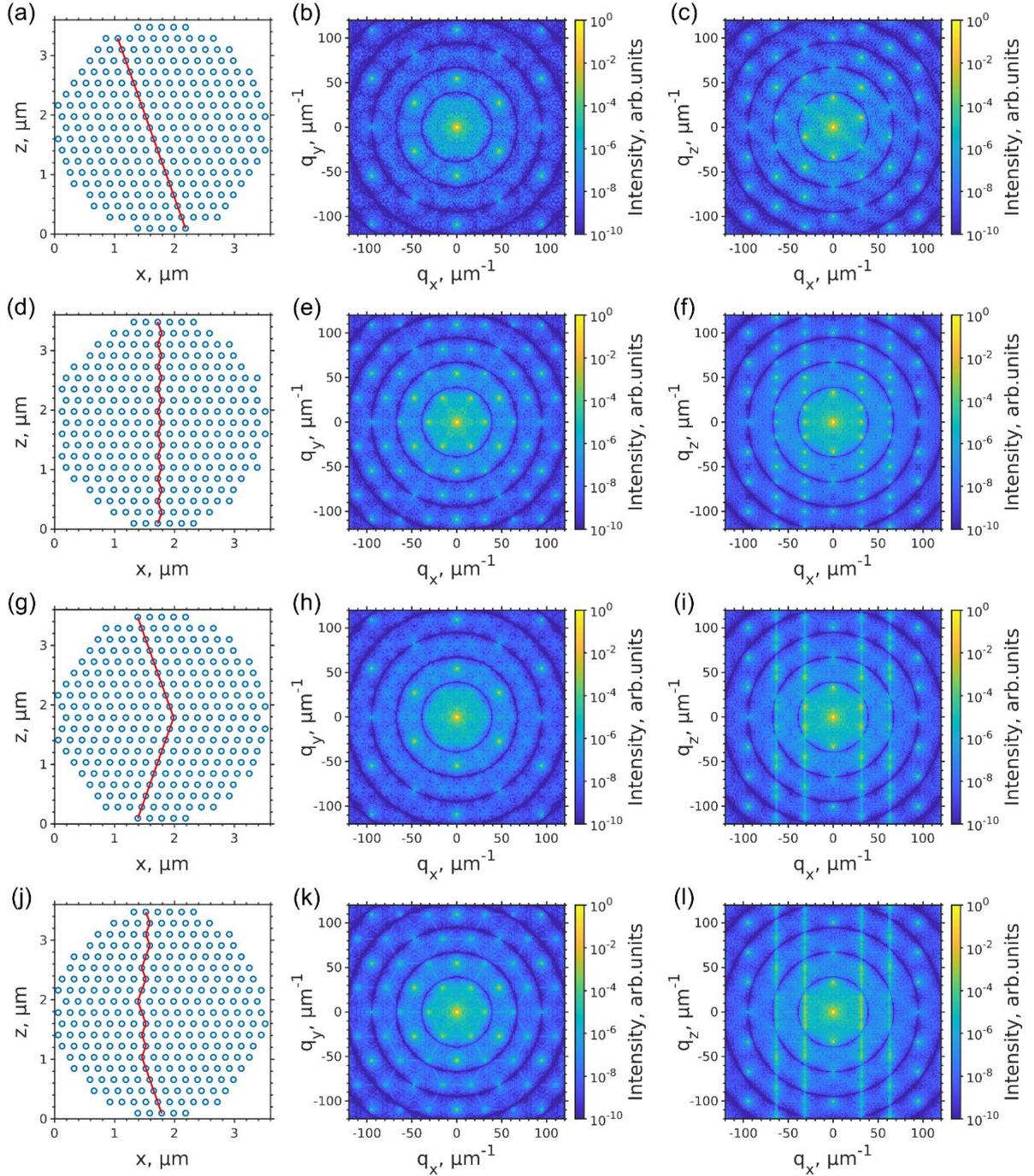

**Figure 4.** Simulation of 2D diffraction patterns from the structures: **(a-c)** *fcc*, **(d-f)** *hcp*, **(g-i)** twinned *fcc* domains, **(j-l)** random *hcp*. The first column contains the simulated structures viewed along $[1\bar{1}0]_{fcc}/[100]_{hcp}$, the stacking direction $[111]_{fcc}/[001]_{hcp}$ is along the *z* axis. The red lines denote the stacking sequence. The second column contains diffraction patterns simulated for an incident beam along the stacking direction $[111]_{fcc}/[001]_{hcp}$. The third

column contains diffraction patterns simulated for an incident beam along the direction $[1\bar{1}0]_{fcc}/[110]_{hcp}$.

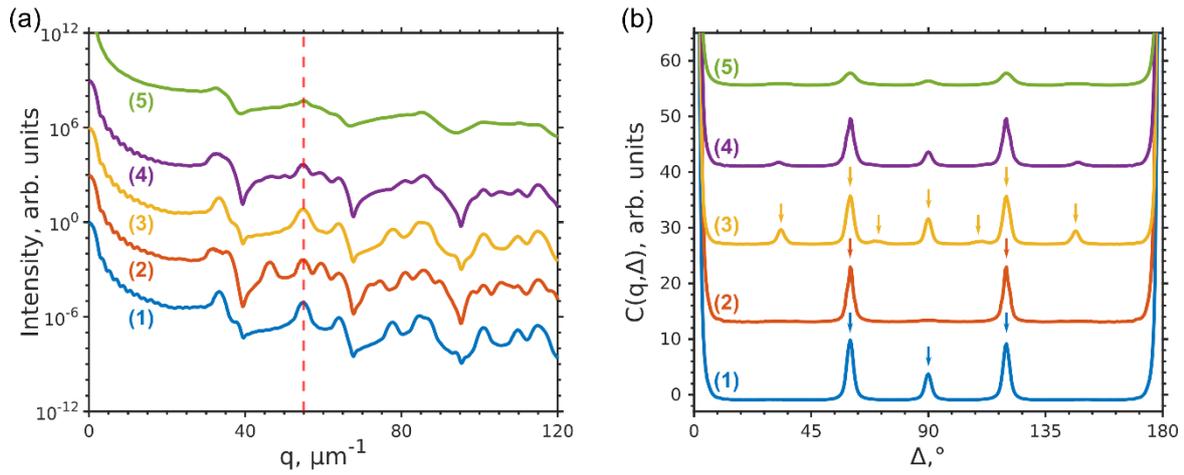

**Figure 5. (a)** Azimuthally averaged values of the 3D intensity distributions simulated for the following structures: *fcc* (1), *hcp* (2), twinned *fcc* domains (3), and random *hcp* (4), and from the experimentally measured sample (5), for comparison. The vertical red dashed line is at $q = 55$ μm$^{-1}$ corresponding to 220$_{fcc}$/110$_{hcp}$ Bragg peaks, for which the CCFs shown in (b) were calculated. **(b)** CCFs $C(q,\varDelta)$ calculated at $q = 55$ μm$^{-1}$ for the simulated 3D diffraction patterns for the following structures: *fcc* (1), *hcp* (2), two twinned *fcc* domains (3), and *rhcp* (4) ), and from the experimentally measured sample (5), for comparison. The arrows show the peak positions calculated for the corresponding structures by a geometrical model.

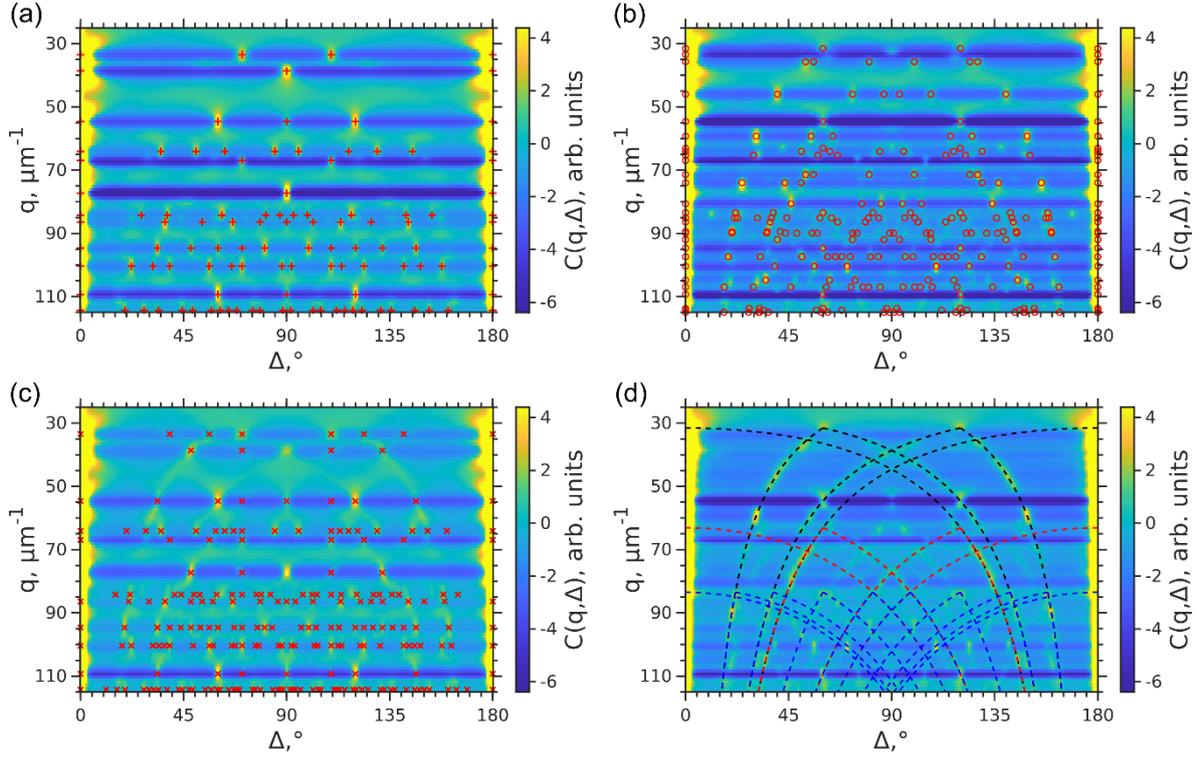

**Figure 6.** Two-dimensional correlation maps $C(q,\Delta)$ calculated in the q-range from 25 nm$^{-1}$ to 115 nm$^{-1}$ for the simulated scattered intensities in 3D from **(a)** *fcc*, **(b)** *hcp*, **(c)** twinned *fcc*, and **(d)** *rhcp* structures. The CCFs are stacked together along the vertical axis $q$. The markers in (a) – (c) indicate the peak positions for the corresponding structures calculated from the geometrical model (see Appendix A). Note that in (c) there are given only the peaks corresponding to the inter-domain correlations between the twin domains. The intra-domain correlations from each domain also give peaks corresponding to an *fcc* structure shown in (a). In (d) the dashed lines indicate the correlations between the Bragg rods. Only correlations within 10*l* (black lines), 20*l* (red lines) and 21*l* (blue lines) Bragg rod families are shown. Correlations between the Bragg rods from different families as well as for higher order families are omitted for clarity.

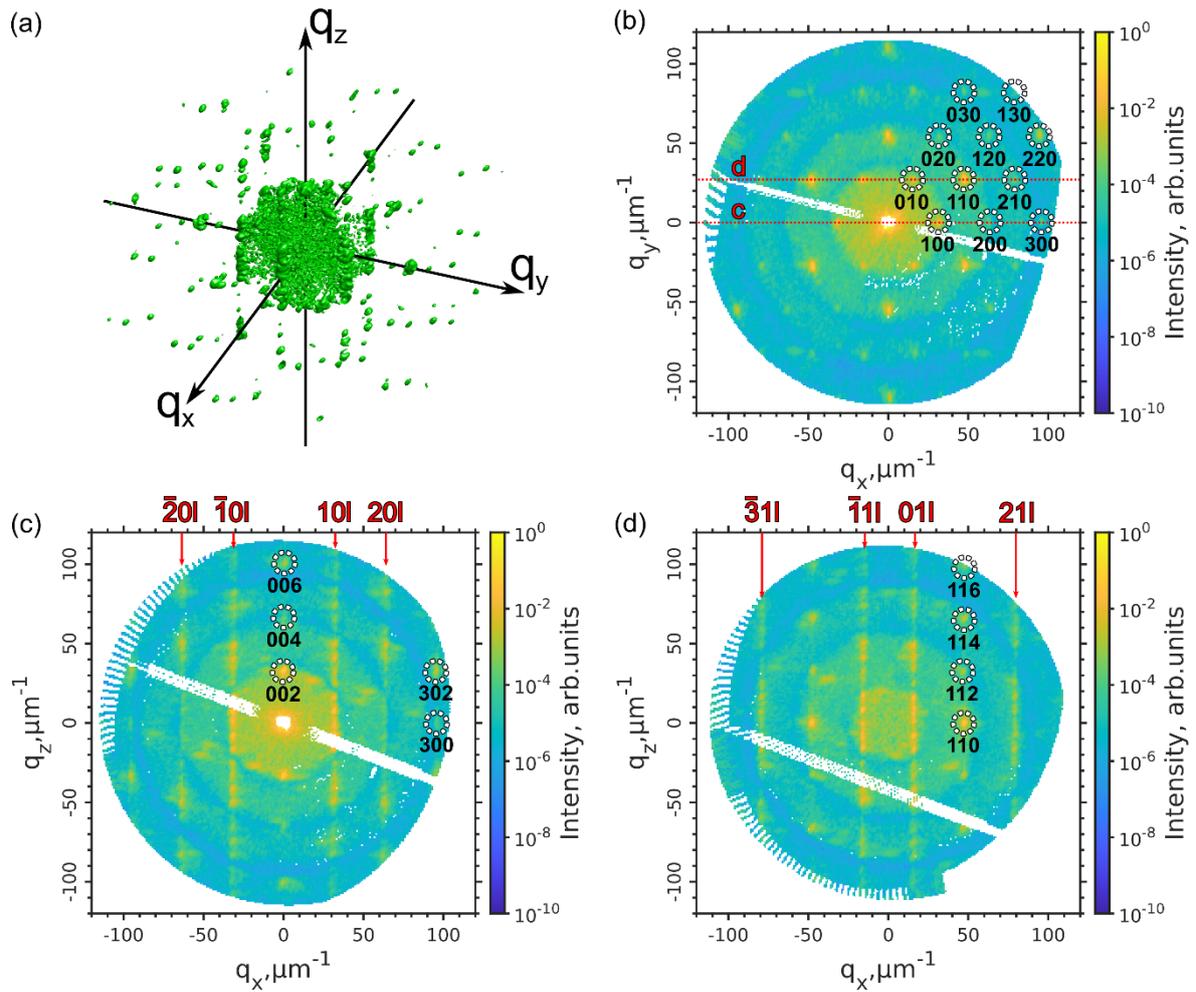

**Figure 7**. **(a)** An isosurface of the measured intensities in 3D reciprocal space. **(b)** A horizontal cut through the origin of reciprocal space. The Bragg peaks are attributed to an hcp lattice. The red lines show the cuts in the panels (c) and (d). **(c)** A vertical cut through the 100 and $\bar{1}00$ reflections, and the origin of reciprocal space. **(d)** A vertical cut through the 010 and $\bar{1}10$ reflections with an offset of 30.5 nm$^{-1}$ along $q_y$ from the origin of reciprocal space. The Bragg rods connecting the Bragg peaks of the 10$l$, 20$l$ and 21$l$ families are indicated with red arrows.

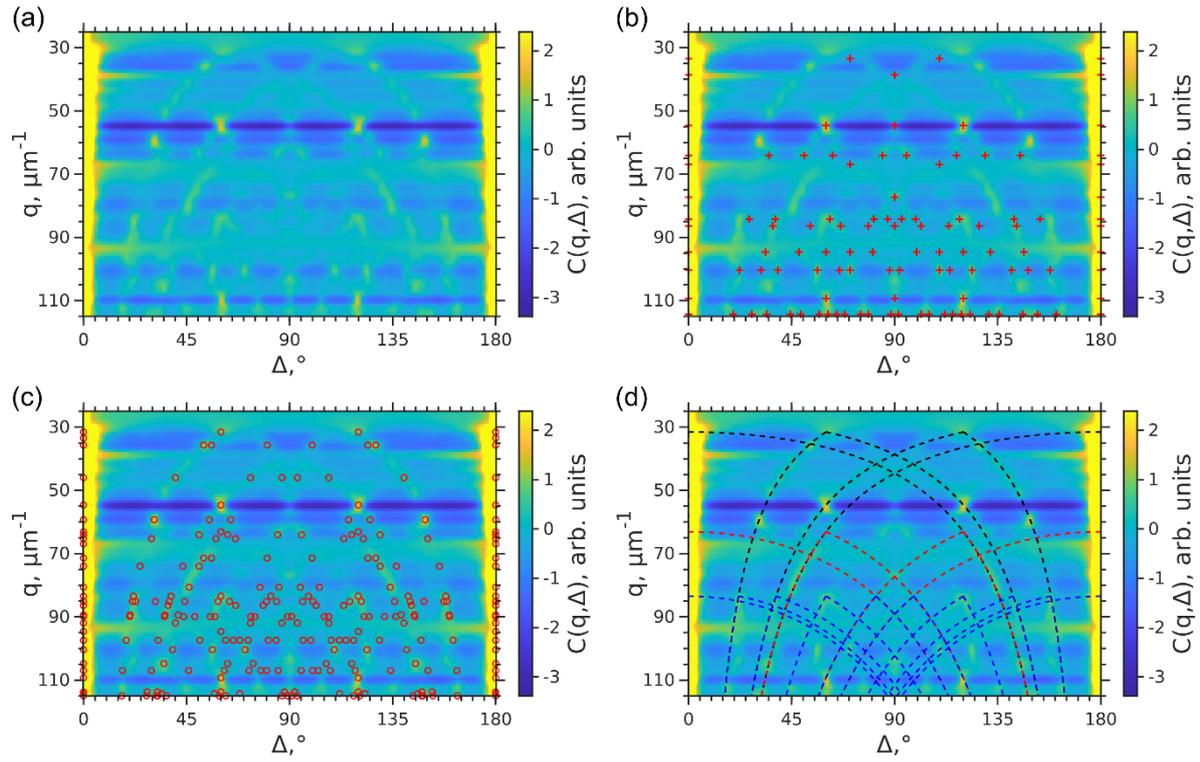

**Figure 8. (a-d)** Two-dimensional correlation maps $C(q,\Delta)$ for the experimentally measured intensity distribution in 3D reciprocal space. (a) Initial correlation map. The markers in (b) – (c) indicate the peaks positions for (b) *fcc* and (c) *hcp* calculated from the geometrical model. In (d) the dashed lines indicate the correlations between the Bragg rods simulated for the *rhcp* structure. Only correlations within 10*l* (black lines), 20*l* (red lines) and 21*l* (blue lines) Bragg rod families are shown. Correlations between the Bragg rods from different families as well as for higher order families are omitted for clarity.

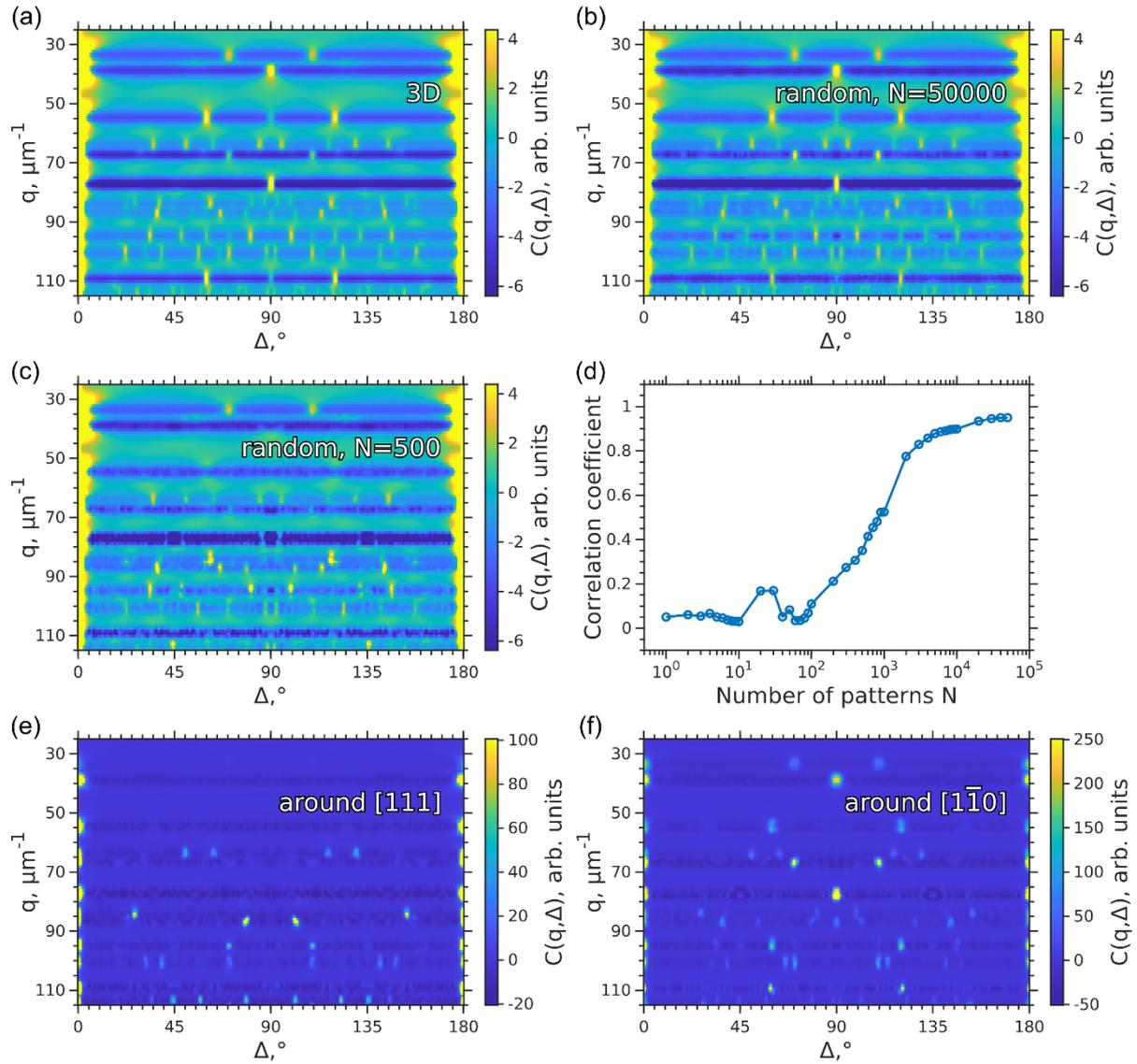

**Figure 9.** Two-dimensional correlation maps $C(q,\Delta)$ calculated for **(a)** the simulated scattered intensities in 3D reciprocal space and **(b) – (e)** 2D diffraction patterns from randomly oriented sample averaged over $5 \cdot 10^4$ **(b)** and $5 \cdot 10^2$ **(c)** patterns. The scattered intensity distribution in 3D reciprocal space and the 2D diffraction patterns were simulated for the same colloidal crystal with an *fcc* structure. **(d)** Pearson correlation coefficient between the CCF maps averaged over different numbers of 2D diffraction patterns and the CCF map calculated for the intensity distribution in 3D reciprocal space for the same sample. **(e) – (f)** Two-dimensional correlation maps $C(q,\Delta)$ calculated for 2D diffraction patterns obtained by rotation of the sample with an *fcc* structure around the $[1\bar{1}1]_{fcc}$ **(e)** and $[1\bar{1}0]_{fcc}$ **(f)** axes.